\documentclass[runningheads,a4paper]{llncs}

\usepackage{amssymb, amsmath}
\usepackage{graphicx}
\usepackage[boxed]{algorithm2e}
\usepackage{tikz}
\usetikzlibrary{arrows, decorations.markings}
\usetikzlibrary{fit}					
\usetikzlibrary{backgrounds}

\newcommand\blackslug{\hbox{\hskip 1pt \vrule width 4pt height 8pt depth 1.5pt
        \hskip 1pt}}
\newcommand\bbox{\hfill \quad \blackslug \medbreak}

\def\c{\hbox{-}\cdots\hbox{-}}

\newcommand{\Proof}{\noindent{\bf Proof.}\ \ }

\newtheorem{main_definitions}{Definition}[section]
\newtheorem{main_corollaries}{Corollary}[section]

\tikzstyle{every node}=[circle, draw, fill=black!50, inner sep=0pt, minimum width=20pt]

\tikzstyle{input}=[circle,
                                    thick,
                                    minimum size=2.2cm,
                                    draw=black!80,
                                    fill=white!20]

\tikzstyle{input2}=[circle,
                                    thick,
                                    minimum size=1.0cm,
                                    draw=black!80,
                                    fill=white!20]

\tikzstyle{matrx}=[rectangle,
                                    thick,
                                    minimum size=1.8cm,
                                    draw=black!80,
                                    fill=white!20]

\tikzstyle{matrx2}=[rectangle,
                                    thick,
                                    minimum size=1.5cm,
                                    draw=black!80,
                                    fill=gray!20]

\tikzstyle{vecArrow} = [thick, decoration={markings,mark=at position
   1 with {\arrow[semithick]{open triangle 60}}},
   double distance=1.8pt, shorten >= 5.5pt,
   preaction = {decorate},
   postaction = {draw,line width=1.4pt, white,shorten >= 4.5pt}]
\tikzstyle{innerWhite} = [semithick, white,line width=1.4pt, shorten >= 4.5pt]

\tikzstyle{background}=[rectangle,
                                                fill=gray!10,
                                                inner sep=0.2cm,
                                                rounded corners=5mm]

\begin{document}

\title{An $\tilde{O}(\frac{1}{\sqrt{T}})$-error online algorithm for retrieving heavily perturbated statistical databases in the low-dimensional querying model}

\author{Krzysztof Choromanski, Afshin Rostamizadeh, Umar Syed}

\institute{Google Research, New York NY 10011, USA}

\maketitle

\begin{abstract}
We give the first $\tilde{O}(\frac{1}{\sqrt{T}})$-error online algorithm for
reconstructing noisy statistical databases, where $T$ is the number of (online)
sample queries received.  The algorithm, which requires only $O(\log T)$
memory, aims to learn a hidden database-vector $w^{*} \in \mathbb{R}^{D}$ in
order to accurately answer a stream of queries regarding the hidden database,
which arrive in an online fashion from some unknown distribution $\mathcal{D}$.
We assume the distribution $\mathcal{D}$ is defined on the neighborhood of a
low-dimensional manifold.  The presented algorithm runs in $O(dD)$-time per
query, where $d$ is the dimensionality of the query-space.  Contrary to the
classical setting, there is no separate training set that is used by the
algorithm to learn the database --- the stream on which the algorithm will be
evaluated must also be used to learn the database-vector. The algorithm only
has access to a binary oracle $\mathcal{O}$ that answers whether a particular
linear function of the database-vector plus random noise is larger than a threshold,
which is specified by the algorithm. We note that we allow for a significant
$O(D)$ amount of noise to be added while other works focused on the low noise $o(\sqrt{D})$-setting.  
For a stream of $T$ queries our algorithm
achieves an average error $\tilde{O}(\frac{1}{\sqrt{T}})$  by filtering out random
noise, adapting threshold values given to the oracle based on its previous
answers and, as a consequence, recovering with high precision a projection of a
database-vector $w^{*}$ onto the manifold defining the query-space. \\
\end{abstract}

\section{Introduction}

Protecting databases that contain sensitive information has become increasingly
important due to its crucial practical applications, such as the 
disclosure of sensitive health data. Privacy preservation 
plays a key role in this setting since such data is often
published in anonymized form so it can be used by analysts and researchers.
Several mechanisms have been proposed, such as differential privacy, that allow
for learning from a database while preserving privacy guarantees
(\cite{D06,Dw10,DNPR10,DNPRY10,NST10}). At the other extreme are many
results showing how database privacy can be compromised by an adversary who is
able to collect perturbated answers to a large number of queries regarding the
database (\cite{Y10,DMT07,DN03,DY08,Chor12}). Existing results related
to breaking the privacy of a database have several key limitations.  For
example, most assume that each query is represented by a vector $q$ of $D$
independent entries taken from some fixed distribution (such as the Gaussian
distribution or a specific discrete distribution), and that this structure is
known to the privacy-breaking algorithm.  Also, most methods learn an
approximation of the unknown database-vector $w^{*}$ that has $L_{2}$ error $\epsilon
D$ for some small constant $\epsilon>0$. Such precision is not
sufficient to obtain $o(1)$-error on the stream of $T$ queries for $T \gg D$,
as is the case in our model.  Further, the focus has typically been on the
offline setting, where the adversary first collects all the queries, then
applies some privacy-breaking algorithm, and finally uses the reconstructed
database-vector to compute good approximations of the statistics he needs. From
the machine learning point of view this means that the overall protocol for the
adversary consists of two distinct phases: a training phase and a testing
phase.  Finally, the memory resources used by privacy-breaking algorithms are
typically not analyzed, even though this is a crucial issue for the setting
considered here, where the number of all the queries $q$ coming in the stream
may be huge. 

The goal of this paper is to present and analyze a database privacy-breaking
algorithm for a more realistic setting in which the limitations described above
are lifted. The entries of the query-vector are not necessarily independent. The
distribution $\mathcal{D}$ of the query-vector is not known to the
adversary.  The adversary is not able to first learn the database-vector
before being evaluated.  Our algorithm uses only $O(\log(T))$-size memory to
process the entire stream of $T$ queries and therefore is well-suited to the
limited resources scenario. To make life of the adversary even more difficult,
we assume that the database mechanism provides only a binary oracle
$\mathcal{O}$ that answers whether the perturbated value of a 
dot-product between the database-vector $w^{*}$ and the query-vector $q$
is greater than a threshold that is specified by the adversary.  Thus the
algorithm has very limited access to the database even in the noiseless
scenario.  Dot-products between query-vector and a database-vector are
considered in most of the settings analyzing database privacy-breaking 
algorithms. Considering this more challenging setting, we will show that much less than
the noisy answer is needed to carry out an effective attack and compromise data privacy.

In some of the mentioned papers an effort is made to learn a good approximation
of the database vector with a small number of queries that is only linear in the size of the database $D$.
We use many more queries but our task is more challenging - we need much more accurate approximation,
and get the information only about the sign of the perturbated product as opposed to the perturbated product itself.
Finally, we are penalized whenever we are making a mistake. Our goal is to minimize the average error
of the algorithm over a long sequence of queries so we need to learn this more accurate approximation very fast.
\\


In this paper we present the first online algorithm that an adversary can use
to reconstruct a noisy statistical database protected by a binary oracle
$\mathcal{O}$ that achieves average error $\tilde{O}(\frac{1}{\sqrt{T}})$ on
the stream of $T$ queries and operates in logarithmic memory. From now on we
will call this algorithm a \textit{learning algorithm}.  The learning algorithm
is given a set of queries taken from some unknown distribution $\mathcal{D}$
defined on a neighborhood of the low-dimensional manifold that it needs to answer in the order that
they arrive (note that the entries of a fixed query do not have to be
independent).  The learning algorithm can use the information learned from
previously collected queries but cannot wait for other queries to learn a more
accurate answer.  Every received query can be used only once to communicate
with a database.  The database mechanism calculates a perturbated answer to the
query and passes the result to the binary oracle $\mathcal{O}$. The binary
oracle uses the threshold provided by the adversary and passes a ``Yes/No"-answer
to him. The error made for a single query is defined as: $|z^{t}-w^{*} \cdot
q^{t}|$, where $q^{t}$ and $z^{t}$ are the query and answer, respectively, 
provided by the learning algorithm in round $t$.  As a byproduct of our methods, we
recover with high precision the projection of the database-vector $w^{*}$ onto
the query-space. Our approximation is within $\tilde{O}(\frac{1}{\sqrt{T}})$
$L_{2}$-distance from the exact projection. By comparison, most of the previous
papers focused on approximating/recovering all but at most a constant fraction
$\epsilon D$ of all the entries of $w^{*}$ which is unacceptably inaccurate in
our learning setting where $T \gg D$.  The assumption that queries are taken
from a low-dimensional manifold is in perfect agreement with recent development
in machine learning (see: \cite{DasguptaF08}, \cite{BelkinN03},
\cite{BaraniukCW10}). It leads to the conclusion that, as stated in \cite{DasguptaF08}:
``a lot of data which superficially lie in a very high-dimensional space
$\mathbb{R}^{D}$ actually have low \textit{intristic dimensionality}, in the
sense of lying close to a manifold of dimension $d \ll D$".  Assume that the
queries are taken from a truly high-dimensional space. Then as long as the number
of all queries is polynomial in $D$, the average distances between them are
substantial.  In this scenario any nontrivial noisy setting prevents the
adversary from learning anything about the database since a single perturbated
answer does not give much information and the probability that a close enough
query will be asked in the future is negligible in $D$. In practice we observe
however that noise can be very often filtered out and a significant number of queries can
give nontrivial information about a database-vector $w^{*}$. In this paper we
explain this phenomenon from the theoretical point of view. Our algorithm
accurately reconstructs the part of the database that regards the
lower-dimensional space used for querying.  We show that this suffices to
achieve average $\tilde{O}(\frac{1}{\sqrt{T}})$-error on the set of $T$ given
queries.
In our model, the number of queries significantly exceeds the dimensionality of the database, and therefore we focus on optimizing our algorithm's time complexity and accuracy as a function of $T$. 
Having said that, in most of the formulas derived in the paper we will also explicitly give the dependence on other parameters of the model such as
the dimensionality of the database $D$ and the dimensionality of the query-space $d$.
We are mainly interested in the setting: $d \ll D \ll T$.
If we use the $O$-notation, where the dependency is not explicitly given then we treat all missing parameters as constants.

It should be also emphasized that, contrary to most previous work on reconstructing databases based on the perturbated statistics, the proposed algorithm does not use
linear programming and thus gives better theoretical guarantees regarding running time than most existing methods. The algorithm uses
a subroutine whose goal is to solve a linear program, however we show this program has a closed-form solution. Therefore we do not need to use any techniques such as
simplex or the ellipsoid method. 
The algorithm is very fast: it needs only $O(dD)$-time per query.
More detailed analysis of the running time of the algorithm as well as memory usage will be given in the Appendix.

\section{Model description and main result}

We will now describe in detail our database access model. 
We assume that the database can be encoded by the database-vector $w^{*} \in \mathbb{R}^{D}$. For definiteness we will consider: $w^{*}_{i} \in [0,1]$ for $i=1,\ldots,D$. 
Our method can be however used in the much more general setting, as long as $w^{*}$ is taken from some fixed ball in $L_{\infty}$. Each query can be represented as a vector $q=(q_{1},\ldots,q_{D})$, 
where: $0 \leq q_{i}\leq 1$ and $q_{1}^{2}+\cdots+q_{D}^{2}>0$.
Queries are taken independently at random from the unknown distribution $\mathcal{D}$ (notice that entries of a fixed query do not have to be independent). 
The distribution $\mathcal{D}$ is defined on some $d$-dimensional linear subspace
$\mathcal{U} \in \mathbb{R}^{d}$ $(d < D)$.
The exact answer to the query is given as $a = \sum_{i=1}^{D} w^{*}_{i}q_{i}$. For the $t^{th}$ coming query $q^{t}$ the learning algorithm $\mathcal{L}$ 
selects the threshold value
$\theta^{t}$ and passes $q^{t}$ to the database mechanism $\mathcal{M}$ which computes $a^{t} = w^{*} \cdot q^{t}$.
The noisy version $\tilde{a}^{t}$ of $a^{t}$ as well as $\theta^{t}$ is passed by $\mathcal{M}$ and $\mathcal{L}$ to the binary oracle $\mathcal{O}$:

$$
 \mathcal{O}(\tilde{a}^{t},\theta^{t}) = \left\{ 
  \begin{array}{l l}
    1 & \quad \text{if $\tilde{a}^{t} > \theta^{t}$,}\\
    0 & \quad \text{otherwise.}
  \end{array} \right.
$$

The value $\mathcal{O}(\tilde{a}^{t},\theta^{t})$ is then given to $\mathcal{L}$. The learner records this value and can also use the information
obtained from previously received queries to give an answer $z^{t}$ to the query $q^{t}$.  However it has only $O(\log(T))$-memory available.
Further, for a fixed query the learner only has one-time access to the binary oracle $\mathcal{O}$.

The noise $\epsilon^{t} = \tilde{a}^{t} - a^{t}$ is generated independently at random and is of the form $D \mathcal{E}$, where $\mathcal{E}$
is some known distribution producing values from some bounded range $[-u,u]$. 
The boundedness assumption is not crucial. Technically speaking,
as long as the random variable is not heavy-tailed (which is a standard assumption), our approach works. 
In fact even this condition is unnecessarily strong.
This will become obvious later when we describe and analyze our method.

This setting covers standard scenarios where computing every single product in the sum of $d$ terms for $w^{*} \cdot q^{t}$
gives an independent bounded error.
We should notice here that in most of the previous papers the magnitude of the noise added was of the order $o(\sqrt{D})$
(see: \cite{Y10,DMT07,DN03,DY08,Chor12}). 
For instance, in \cite{DMT07} the authors reconstruct a database that agrees with the groundtruth one on all but $(2c\alpha)^{2}$ entries,
where $\alpha$ is a noise magnitude and $c>0$ is a constant.
Thus, even though previous works do not assume that noise was added independently
for every query, the average error per single product in the dot-product sum was only of the magnitude $o(\frac{1}{\sqrt{D}})$.
This assumption significantly narrows the range of possible applications. This is no longer the case in our setting, where some
mild and reasonable assumptions regarding independence of noise added to different queries and low-dimensionality of querying space leads to a model much more robust to noise.  
We will assume that $\epsilon^{t}$ do not have singularities, i.e.
$\mathbb{P}(\epsilon^{t}=c) = 0$ for any fixed $c$.

We need a few more definitions.

\begin{definition}
 We say that a vector $w$ computed by the learning algorithm $\epsilon$-approximates database-vector $w^{*}$ if
 $|\Pi_{\mathcal{U}}(w)-\Pi_{\mathcal{U}}(w^{*})|_{\infty} \leq \epsilon$, where $\Pi_{\mathcal{U}}(v)$ stands for the projection of $v$ onto $d$-dimensional querying space $\mathcal{U}$.
\end{definition}

\begin{definition}
Let $\mathcal{Q}$ be a probability distribution on the unit sphere $\mathcal{S}(0,1)$ in $L_{2}$.
For a fixed vector $q \in \mathcal{S}(0,1)$ we denote by $p^{\mathcal{Q}}_{q,\theta}$ the probability that a vector
$x$ selected according to $\mathcal{Q}$ satisfies: $q \cdot x \geq \cos(\theta)$. 
\end{definition}

\begin{definition}
Take a distribution $\mathcal{D}$ from which queries are taken.
Assume that $\mathcal{D}$ is defined on the $d$-dimensional space $\mathcal{U}$ with orthonormal basis $\mathcal{B}$.
Denote by $\mathcal{D}_{n}$ the normalized version of $\mathcal{D}$ and by $\mathcal{B}_{n}$ the normalized version of $\mathcal{B}$ 
(all vectors rescaled to length $1$ in the $L_{2}$-norm). 
Then we define: $p_{\mathcal{D},\theta} =  \min_{q \in \mathcal{B}_{n}} (p^{\mathcal{D}_{n}}_{q, \theta})$.
\end{definition}

The error $\epsilon_{q}$ the algorithm is making on each query $q$ is defined as the absolute value of the difference between the exact answer to the query
and the answer that is provided by the algorithm. The average error on the set of queries: $q^{1},...,q^{T}$ is defined as $\epsilon_{av} = \frac{1}{T}\sum_{i=1}^{T} \epsilon_{q^{i}}$.
Let us state now main result of this paper.

\begin{theorem}
\label{core_thm}
Let $q^{1},\ldots,q^{T}$ be a stream of query-vectors coming in an online fashion from some $d$-dimensional subspace, where:
$0 \leq q^{t}_{i} \leq 1$ for $i=1,\ldots,d$ and each $q^{t}$ is a nonzero vector. Then there exists an algorithm
$\mathcal{A}$lg using $O(\log(T))$-memory, acting according to the protocol defined above, 
and achieving average error:
$$e_{av} = O(\frac{1}{\sqrt{T}}(rD^{\frac{7}{2}}d+\sqrt{D}\log(T)))$$ with probability $p_{succ} \geq 1 - O(\frac{\log(dDT)}{T^{3}} + \frac{d\log(dT)}{T^{30}})$,
where $r=\frac{2}{p_{\mathcal{D},\phi}^{2}}$ and $\phi = 2\arcsin(\frac{1}{64\sqrt{d}})$. 
\end{theorem}

We will give this algorithm, called \textit{OnlineBisection} algorithm, in the next section. 
Notice that $\phi$ is well approximated by $\frac{1}{32\sqrt{d}}$. To see what the magnitude of $r$ is
in the worst-case scenario it suffices to analyze the setting where $q$ is chosen uniformly at random from the query-space
$\mathcal{U}$.

If this is the case then one can notice that $p_{\mathcal{D},\phi}$ is of the order $\Omega(2^{-d\log(d)})$ thus $r = O(2^{d\log(d)})$. If however there exists a basis of $\mathcal{U}$ such that most of the mass of $\mathcal{D}$
is concentrated around vectors from the basis then standard analysis leads to the $\frac{1}{poly(d)}$-lower bound on $p$,
i.e. $poly(d)$-upper bound on $r$ (where $poly(d)$ is a polynomial function of $d$).
 
Theorem \ref{core_thm} implies a corollary regarding the batch version of the algorithm, where test and training set are clearly separated (the proof of that corollary will be given in the Appendix):

\begin{corollary}
\label{batch_corollary}
  Let $w_T$ denote the final hypothesis constructed by the
  \textit{OnlineBisection} algorithm after consuming $T$ queries drawn
  from an unknown distribution $\mathcal{D}$. Then the following
  inequality holds with probability at least $1 - O \Big(
  \frac{\log(dDT)}{T^3} + \frac{d \log(dT)}{T^{30}} \Big)$ for any future
  queries $q$ drawn from $\mathcal{D}$:
  \begin{equation*}
    \mathrm{E}_{q \sim \mathcal{D}} \big[ | w_T \cdot q - w^* \cdot q | \big]
    \leq \frac{\sqrt{D} \log(T)}{\sqrt{T}} \,.
  \end{equation*}
\end{corollary}

In the subsequent sections we will prove Theorem \ref{core_thm}
and conduct further analysis of the algorithm.
Unless stated otherwise, $\log$ denotes the natural logarithm.

\begin{algorithm}[H]
 \textbf{Algorithm 1 - \textit{OnlineBisection}} \\
 \textbf{Input:} Stream $q^{1},\ldots,q^{T}$ of $T$ queries, database mechanism $\mathcal{M}$ and binary oracle $\mathcal{O}$.\\
 \textbf{Output:} A sequence of answers $(w^{1} \cdot q^{1},\ldots,w^{T} \cdot q^{t})$, returned online.\\
 \Begin{
  Choose an orthonormal basis $\mathcal{C}=\{e^{1},\ldots,e^{d}\}$ of $\mathcal{U}$.\\
  Let $\phi = 2\arcsin(\frac{1}{64\sqrt{d}})$.\\
  Let $\mathcal{I}_{i} = [-\sqrt{D},\sqrt{D}]$, $N^{+}_{i} = 0$ and $N^{-}_{i}=0$ for $i=1,\ldots,d$.\\
  \For{$t = 1, \ldots, T$} {
    Output $w_{approx} \cdot q^t$ for any $w_{approx} = f_{1}e^{1}+\cdots+f_{d}e^{d}$, where $f_{i} \in \mathcal{I}_{i}$, $i=1,\ldots,d$.\\
    \textbf{if} $|\mathcal{I}_{i}| \leq \frac{\log(T)}{\sqrt{Td}}$ for $i=1,...,d$ \textbf{continue}. \\ 
    \If{$\exists i^{*} \in \{1,\ldots,d\}$ such that $\arccos(e^{i^{*}},\frac{q^t}{\|q^t\|_{2}}) \leq \phi$} {
      Let $m=\max_{f_{1} \in \mathcal{I}_{1},\ldots,f_{d} \in \mathcal{I}_{d}}\sum_{i=1}^{d} f_{i} e^{i} \cdot (-q^t)$.\\
      Let $M=\max_{f_{1} \in \mathcal{I}_{1},\ldots,f_{d} \in \mathcal{I}_{d}}\sum_{i=1}^{d} f_{i} e^{i} \cdot q^t$.\\   
      Let $b = \mathcal{O}(\mathcal{M}(q^t), \frac{m+M}{2})$.\\
      If $b>0$ update $N^{+}_{i^{*}} \leftarrow N^{+}_{i^{*}} + 1$, otherwise update $N^{-}_{i^{*}} \leftarrow N^{-}_{i^{*}} + 1$.\\
    }
    Let $\Delta p = \mathbb{P}(-\frac{|\mathcal{I}_{1}|}{8D} \leq \mathcal{E} \leq \frac{|\mathcal{I}_{1}|}{8D})$, $N_{i} = N^{+}_{i} + N^{-}_{i}$ and $N_{crit} = \frac{30 \log(T)}{\Delta p^{2}}$.\\
    \If{$N_{i} \geq N_{crit}$ for $i=1,\ldots,d$} {
      Run \textit{ShrinkHyperCube}($\mathcal{I}_{1},\ldots,\mathcal{I}_{d}$, $N^{+}_{1},\ldots,N^{+}_{d}$, 
                          $N^{-}_{1},\ldots,N^{-}_{d}$).\\
      Update: $N^{+}_{i} \leftarrow 0, N^{-}_{i} \leftarrow 0$ for $i=1,\ldots,d$.\\                    
    }
  }
 }
\end{algorithm}

\section{The Algorithm}
We will now present an algorithm (Algorithm 1) that achieves theoretical
guarantees from Theorem \ref{core_thm}.  Our algorithm, called
\textit{OnlineBisection}, maintains a tuple of intervals
$(\mathcal{I}_{1},\ldots,\mathcal{I}_{d})$ which encode a hypercube that
contains the database-vector $w^{*}$ (projected onto $\mathcal{U}$) with very
high probability. For each coming query-vector $q^t$ the algorithm outputs an
answer $w_{approx} \cdot q^t$, where $w_{approx}$ is an arbitrarily selected
vector in the current hypercube. The query-vectors received by the algorithm
are used to progressively shrink the hypercube. 

As the hypercube shrinks,
vector $w_{approx}$ $\epsilon$-approximates $w^{*}$ for smaller values of $\epsilon$.
When the hypercube is large the errors made by the algorithm will be large, but
on the other hand larger hypercubes are easier to shrink since they require
fewer queries to ensure that hypercube continues to contain $w^*$ (with very
high probability) after shrinking. This observation plays a crucial role in
establishing upper bounds on the average error made by the algorithm on the
sequence of $T$ queries.

After outputting an answer for query-vector $q^t$, the algorithm checks whether
$q^t$ has a large inner product with at least one vector in an orthonormal
basis $\mathcal{C}=\{e^{1},\ldots,e^{d}\}$ of $\mathcal{U}$.  If so, $q^t$
represents an observation for that basis vector; whether it is a positive or
negative observation depends on the response of the binary oracle
$\mathcal{O}$. The threshold given by the algorithm to $\mathcal{O}$ is chosen
by solving the linear program $\max_{y \in \mathcal{H}\mathcal{C}} q \cdot y$
for $q = q^t$ and $q = -q^t$, where $\mathcal{H}\mathcal{C}$ is the current
hypercube.  As
we will see in Section \ref{sec:running}, this linear program is simple enough
that there is a closed-form expression for its optimal value.  So we do not
need to use the simplex method or any other linear programming tools.

\begin{algorithm}[H]
 \textbf{Algorithm 2 - \textit{ShrinkHyperCube}} \\
 \textbf{Input:} $\mathcal{I}_{1}=[x_{1},y_{1}],\ldots,\mathcal{I}_{d}=[x_{d},y_{d}]$, $N^{+}_{1},\ldots,N^{+}_{d}$, $N^{-}_{1},\ldots,N^{-}_{d}$.\\
 \textbf{Output:} Updated hypercube $(\mathcal{I}_{1},\ldots,\mathcal{I}_{d})$.\\
 \Begin{
   Let $\alpha = \frac{3}{4}$, 
      $\Delta p = \mathbb{P}(-\frac{|\mathcal{I}_{1}|}{8D} \leq \mathcal{E} \leq \frac{|\mathcal{I}_{1}|}{8D})$, 
              $p_{1} = \mathbb{P}(\mathcal{E}>\frac{|\mathcal{I}_{1}|}{8D})$ and $N_i = N^{+}_i + N^{-}_i$.\\
   \For{$i=1,\ldots,d$} {
     \eIf{$N^{+}_{i} > N_i p_1+ \frac{N_i \Delta p}{2}$}{$\mathcal{I}_{i} \leftarrow [y_{i} - \alpha(y_{i}-x_{i}),y_{i}]$\;}
     {$\mathcal{I}_{i} \leftarrow [x_{i},x_{i} + \alpha(y_{i}-x_{i})]$\;}
   }
 }
\end{algorithm}

The optimal values $m$ and $M$ of the linear programs solved by the
\textit{OnlineBisection} algorithm represent the smallest and largest possible
value of the inner product of the query-vector and a vector from the current
hypercube.  The true value lies in the interval $[m,M]$. By choosing the
average of these two values as a threshold for the oracle we are able to
effectively shrink direction $i^{*}$.  The intuition is that if the 
query-vector forms an angle $\alpha=0$ with this direction and there is no noise
added then by choosing the average we basically perform standard binary search
for $q$. Since $\alpha$ is not necessarily $0$ but is relatively small (and
noise is added that perturbates the output), the search is not exactly binary.
Instead of two disjoint subintervals of $I_{i^{*}}$ we get two intervals whose
union is $I_{i^{*}}$ but that intersect.  Still, each of them is only of a
fraction of the length of $I_{i^{*}}$ and that still enables us to
significantly shrink each dimension whenever a sufficient number of
observations have been collected for each basis vector --- specifically,
$N_{crit}$ observations --- by calling the $\textit{ShrinkHyperCube}$
subroutine (Algorithm 2). 

Every shrinking of the hypercube decreases each edge by a factor $\alpha$ for
some $0 < \alpha < 1$.  A logarithmic number of shrinkings is needed to ensure
that any choice of $w_{approx}$ in the hypercube will give an error of the
order $\tilde{O}(\frac{1}{\sqrt{T}})$. Notice that $N_{crit}$ grows with $T$,
which reflects the fact that for smaller hypercubes more observations are
needed to further shrink the hypercube while preserving the property that it
contains the database-vector $w^{*}$ with very high probability.  
This is the case since if the hypercube is small we already know a good approximation of the database
vector so it is harder to find even more accurate one under the same level of noise.
When the
hypercube is small enough (condition: $|\mathcal{I}_{i}| \leq
\frac{\log(T)}{\sqrt{Td}}$ for $i=1,...,d$) there is no need to shrink it
anymore since each vector taken from the hypercube is a precise enough
estimate of the database vector.

Note that choosing an orthonormal basis $\mathcal{C} = \{e^{1},\ldots,e^{d}\}$
of $\mathcal{U}$ does not require the knowledge of the distribution
$\mathcal{D}$ from which queries are taken. We only assume that queries are
from a low-dimensional linear subspace $\mathcal{U}$ of $d$ dimensions. It
suffices to have as $\{e^{1},\ldots,e^{d}\}$ some orthonormal basis of that
linear subspace. There are many state-of-the-art mechanisms (such as PCA) that
are able to extract such a basis, and thus we will not focus on that, but
instead assume that such an orthonormal system is already given. Notice that in
practice those techniques should be applied before our algorithm can be run.
Since such a preprocessing phase requires sampling from $\mathcal{D}$ but does
not require an access to the database system, we can think about it as a
preliminary period, where evaluation is not being conducted. 


%

\section{Theoretical analysis}

In this section we prove Theorem \ref{core_thm}.
We start by introducing several technical lemmas.
Their proofs will be given in the Appendix. We prove here how those lemmas can be combined to obtain our main result.

We denote: $h_{T} = \frac{\sqrt{T}}{\log(T)}$. Thus the stopping condition for shrinking the hypercube is of the form:
$|\mathcal{I}_{i}| \leq \frac{1}{\sqrt{d}h_{T}}$ for $i=1,...,d$.

We start with the standard concentration result regarding binomial random variables.
\begin{lemma}
\label{segment_lemma}

%
%
%

Let $Z^{m}=Bin(m,p_{1})$, $W^{m}=Bin(m,p_{1}+\Delta p)$ and $\mu_{1} = m p_{1}$. 

Then the following is true:
\begin{equation}
\mathbb{P}(Z^{m} \geq \mu_{1} + \frac{m \Delta p}{2}) \leq  e^{- \frac{m (\Delta p)^{2}}{10}},
\end{equation}

\begin{equation}
\mathbb{P}(W^{m} < \mu_{1} + \frac{m \Delta p}{2}) \leq e^{- \frac{m (\Delta p)^{2}}{10}}.
\end{equation}

\end{lemma}


\begin{main_definitions}
Let $\mathcal{H}\mathcal{C}$ be a $d$-dimensional hypercube in $\mathbb{R}^{D}$. We denote by $l(\mathcal{H}{\mathcal{C}})$
the length of its side measured according to the $L_{2}$-norm (recall that all the sides of a hypercube have the same length).
\end{main_definitions}

Next lemma is central for finding an upper bound on the average error made by the algorithm. 

\begin{lemma}
\label{cuboid_lemma}
Let $(q_{1},\ldots,q_{T})$ be a sequence of $T$ queries. Let $\mathcal{H}\mathcal{C}_{0},\ldots,\mathcal{H}\mathcal{C}_{s}$
be a sequence of $d$-dimensional hypercubes in $\mathbb{R}^{D}$. 
Assume that $l(\mathcal{H}\mathcal{C}_{i+1}) \leq \alpha l(\mathcal{H}\mathcal{C}_{i})$ 
for $i=0,\ldots,s-1$ and some $0 < \alpha < 1$.
Denote $l(\mathcal{H}\mathcal{C}_{0}) = L \leq D$ and assume that $s = \frac{1}{\log_{2}(\frac{1}{\alpha})}\log_{2}(L\sqrt{d}h(T))$, where $h(T)$ is some function of $T$. 
Assume that $w^{*} \in \mathcal{H}\mathcal{C}_{0} \bigcap \cdots \bigcap \mathcal{H}\mathcal{C}_{s}$.
Let $\mathcal{E}$ be a random variable defined on the interval $[-u,u]$ for some constant $u>0$, with density $\rho$ continuous at $0$, and
such that $\rho(0)>0$. Define $\phi_{\epsilon}(i) = \mathbb{P}(-\frac{L\alpha^{i}(\frac{1}{4}-\epsilon)}{D} < \mathcal{E} \leq \frac{L\alpha^{i}(\frac{1}{4}-\epsilon))}{D}$ 
for some constant $0<\epsilon \leq \frac{1}{8}$.
Let $m_{i} = \frac{1}{\phi_{\epsilon}^{2}(i)}C\log(T)$ for some constant $C>0$ and let
$k_{i} = m_{i} r$ for some other constant $r>0$ and $i=0,\ldots,s$. Assume that learning algorithm uses a vector $w_{approx} \in \mathcal{H}\mathcal{C}_{0}$
to answer first $k_{0}$ queries, a vector $w_{approx} \in \mathcal{H}\mathcal{C}_{1}$ to answer next $k_{1}$ queries, etc. 
Assume also that an algorithm uses a vector $w_{approx} \in \mathcal{H}\mathcal{C}_{s}$ to answer remaining $T - \sum_{i=0}^{s} k_{i}$ queries.
Then the following is true about the cumulative error $\epsilon_{cum}$ made by the algorithm:
$$\epsilon_{cum} = O(L^{2}D^{\frac{5}{2}}dr\log(T)h(T) + \frac{\sqrt{D}T}{h(T)}).$$
\end{lemma}

In the following lemma we analyze cutting the hypercube according to some linear threshold.

\begin{lemma}
\label{cut_lemma}
Let $w \in \mathbb{R}^{D}$, let $\{v^{1},\ldots,v^{d}\}$ be a system of pariwise orthogonal vectors such that $v^{i} \in \mathbb{R}^{D}$, $\|v^{i}\|_{2}=L$ for $i=1,..,d$ and let $\mathcal{H}\mathcal{C}=\{w+ \sum_{i=1}^{d}f_{i}v^{i}:
f_{1},\ldots,f_{d} \in [0,1]\}$ be a $d$-imensional hypercube.
Let $e$ be a unit-length vector in $L_{2}$ that is parallel to $v^{1}$, i.e. $e = \frac{1}{\sqrt{L}}v^{1}$. 
Let $z$ be a unit-length vector satisfying: $z \cdot e \geq \cos(\theta)$ for some $0 < \theta < \frac{\pi}{2}$.
Let $0 < \beta < 1$. Define $m = \min_{y \in \mathcal{H}\mathcal{C}} y \cdot z$ and $M =  \max_{y \in \mathcal{H}\mathcal{C}} y \cdot z$. Let $\mathcal{H}\mathcal{C}_{l} = \{y \in \mathcal{H}\mathcal{C}: z \cdot y \leq m + \beta(M-m)\}$
and $\mathcal{H}\mathcal{C}_{r} = \{y \in \mathcal{H}\mathcal{C}: z \cdot y > m + \beta(M-m)\}$.
Then for $\epsilon = 8\sin(\frac{\theta}{2}) \sqrt{d}$:
\begin{equation}
\label{maxmin_1}
\max_{y \in \mathcal{H}\mathcal{C}_{l}} e \cdot y - \min_{y \in \mathcal{H}\mathcal{C}_{l}} e \cdot y
\leq L(\beta + \epsilon)
\end{equation}

and

\begin{equation}
\label{maxmin_2}
\max_{y \in \mathcal{H}\mathcal{C}_{r}} e \cdot y - \min_{y \in \mathcal{H}\mathcal{C}_{r}} e \cdot y
\leq L(1-\beta + \epsilon).
\end{equation}
\end{lemma}


We are ready to prove Theorem \ref{core_thm} assuming that presented lemmas are true.

\Proof
Let $L=2\sqrt{D}$.
Let us notice that the algorithm can be divided into $s+1$ phases, where 
in the $i^{th}$ phase ($i=0,\ldots,s$) all the intervals $\mathcal{I}_{i}$ are of length $L\alpha^{-i}$
and $s=\frac{1}{\log_{2}(\frac{1}{\alpha})}\log_{2}(L\sqrt{d}h_{T})$.
Indeed, whenever the shrinking is conducted, the length of each side of the hypercube decreases
by a factor $\frac{1}{\alpha}$ (see subroutine \textit{ShrinkHyperCube}), the initial lengths are $2\sqrt{D}$ 
and the shrinking is not performed anymore if the side of each length is at most $\frac{1}{\sqrt{d}h_{T}}$.
We will call those phases: \textit{1st-phase}, \textit{2nd-phase}, etc.
Notice also that the value of the parameter $N_{crit}$ is constant across a fixed phase since this number
changes only when \textit{ShrinkHyperCube} subroutine is performed. Let us denote the value of 
$N_{crit}$ during the $i^{th}$ phase of the algorithm as $n_{i}$.
Notice that $n_{i} = \frac{30 \log(T)}{\Delta p_{i}^{2}}$, where $\Delta p_{i}$ is the value of the parameter
$\Delta p$ of the algorithm used in the $i^{th}$ phase.
Denote by $k_{i}$ the number of queries that need to be processed in the $i^{th}$ phase for $i=0,\ldots,s-1$.
Parameter $k_{i}$ is a random variable but we will show later that with high probability: $k_{i} \leq n_{i} r$ 
for $i=0,\ldots,s-1$, where: $r=\frac{2}{(p_{\mathcal{D},\phi})^{2}}$. Assume now that this is the case.
Denote by $\mathcal{H}\mathcal{C}_{0},\ldots,\mathcal{H}\mathcal{C}_{s}$ the sequence of hypercubes constructed by 
the algorithm. Assume furthermore that $w^{*} \in \mathcal{H}\mathcal{C}_{0} \cap \cdots \cap \mathcal{H}\mathcal{C}_{s}$.
Again, we have not proved it yet, we will show that this happens with high probability later. 
However we will prove now that under these two assumptions we get the
average error proposed in the statement of Theorem \ref{core_thm}. Notice that under these assumptions we can 
use Lemma \ref{cuboid_lemma} with $L=2\sqrt{D}$, $h(T)=h_{T}$, $\phi_{\epsilon}(i) = \Delta p_{i}$, $C=30$, $m_{i}=n_{i}$.
We get the following bound on the cumulative error:
\begin{equation}
\epsilon_{cum} = O(D^{\frac{7}{2}}dr\log(T)h(T)+\frac{\sqrt{d}T}{h(T)}).
\end{equation}
Thus the average error is at most $\epsilon_{av} \leq \frac{\epsilon_{cum}}{T}$. By using the expression
$h(T)=\frac{\log(T)}{\sqrt{T}}$ in the above formula, we obtain the bound from the statement of Theorem \ref{core_thm}.

It remains to prove that our two assumptions are correct with high probability and find a lower bound on this
probability that matches the one from the statement of the theorem. We will do it now.
Let us focus on the $i^{th}$ phase of the algorithm. First we will find an upper bound on the probability that
the number of queries processed in this phase is greater than $k_{i}$.
Fix a vector $e^{j}$ from the orthonormal basis $\mathcal{C}$. The probability that a new query $q$ is within angle
$\phi$ from $e^{j}$ is at least $p=p_{\mathcal{D},\phi}$, by the definition of $p_{\mathcal{D},\phi}$.
Assume that $u_{i}$ queries were constructed. By standard concentration inequalities, such as Azuma's inequality,
we can conclude that with probability at least $1-e^{-2u_{i}(\frac{p}{2})^{2}}$ at least $\frac{u_{i}p}{2}$ of those queries
will be within angle $\phi$ from $e^{j}$. If we take: $u_{i} \geq \frac{2n_{i}}{p}$, then we conclude that with probability
at least $1-e^{-2u_{i}(\frac{p}{2})^{2}}$ at least $n_{i}$ of those queries will be within angle $\phi$ from $e^{j}$.
Denote $u_{i} = n_{i} r$, where $r>\frac{2}{p}$. We see that the considered probability is at least 
$1-e^{-\frac{p^{2}}{2}n_{i}r}$. Using the expression on $n_{i}$ we get that this probability is at least 
$1-e^{-30r\frac{p^{2}}{2}\log(T)}$.
Notice that when $n_{i}$ queries within angle $\phi$ from a given vector $e^{j} \in \mathcal{C}$ are collected,
the $j^{th}$ dimension is ready for shrinking. Thus taking union bound over $O(\log(dT))$ phases and all $d$ dimensions
we see that if we take $k_{i} =rn_{i}$, where: $r= \frac{2}{p^{2}}$, then with probability at most $\frac{d\log(dT)}{T^{30}}$ some
$i^{th}$ phase of the algorithm for $i \in \{0,\ldots,s-1\}$ will require more than $k_{i}$ queries.
Now let us focus again on the fixed $i^{th}$ phase of the algorithm. 
Assume that \textit{ShrinkHyperCube} subroutine is being run. Fix some dimension $j \in \{1,\ldots,d\}$.
We know that, with high probability, at least $n_{i}$ queries $q$ that were within angle $\phi$ from the vector $e^{j} \in \mathcal{C}$
were collected.
Denote by $w^{*}_{j}$ the $j^{th}$ coordinate of $w^{*}$. Let $\mathcal{I}_{j}=[x_{j},y_{j}]$ and assume that $w^{*}_{j} \in [x_{j},y_{j}]$.
Let us assume that the \textit{ShrinkHyperCube} subroutine replaced $\mathcal{I}_{j}=[x_{j},y_{j}]$ by
$\tilde{\mathcal{I}_{j}}$. We want to show that with high probability segment $\tilde{\mathcal{I}_{j}}$ is constructed
in such a way that $w^{*}_{j} \in \tilde{\mathcal{I}_{j}}$. Denote $l=y_{j}-x_{j}$ and $\delta = (\alpha-\frac{1}{2})l$. Notice first that if
$w^{*}_{j} \in [(1-\alpha)l,\alpha l]$ then $w^{*}_{j}$ will be in $\tilde{\mathcal{I}_{j}}$ since no matter how
$\tilde{\mathcal{I}_{j}}$ is constructed, it always contains $[(1-\alpha)l,\alpha l]$. So let us assume that this is not the case.
Thus we have either $w^{*}_{j} \in [x_{j},x_{j} + (1-\alpha)l]$ or $w^{*}_{j} \in [y_{j}-(1-\alpha) l,y_{j}]$.
Let us assume first the former. Consider a query-vector $q$ within angle $\phi$ of $e^{j}$ that contributed to 
$N^{+}_{j}$. Let us denote by $p_{+}$ the probability of the following event $\mathcal{F}_{q}$: for $q$ the oracle $\mathcal{O}$ 
gives answer: ``greater than 0".
Observe that the total error made by the database mechanism $\mathcal{M}$ while computing the dot-product:
$w^{*} \cdot q$ is $D\mathcal{E}$.
Now notice, that by Lemma \ref{cut_lemma} and the definition of $\mathcal{E}$, probability $p_{+}$ 
is at most $\mathbb{P}(D \mathcal{E} > \delta - \epsilon l)$, where: $\epsilon = 8 \sin(\frac{\phi}{2})\sqrt{d}=\frac{1}{8}$. 
Thus we get: $p_{+} \leq \mathbb{P}(\mathcal{E} > \frac{(\alpha-\frac{1}{2}-\epsilon)l}{D})$. Notice that in the $i^{th}$
phase the hypercube under consideration has the side of length exactly $\alpha^{i}$. Thus, since $\alpha=\frac{3}{4}$, we get:
$p_{+} \leq \mathbb{P}(\mathcal{E} > \frac{(\frac{1}{4}-\epsilon)L\alpha^{i}}{D})$.
Let us assume now that $w^{*}_{j} \in [y_{j}-(1-\alpha) l,y_{j}]$. We proceed with the similar analysis as before.
We see that the probability $P_{+}$ of an event $\mathcal{F}_{q}$ is at least $\mathbb{P}(D\mathcal{E} \geq -\delta + \epsilon l)$.
Thus we obtain: $P_{+} \geq \mathbb{P}(\mathcal{E} \geq -\frac{(\frac{1}{4}-\epsilon)L\alpha^{i}}{D})$.
But now we see, by Lemma \ref{segment_lemma}, using: $m = N_{i}$, $p_{1}=\mathbb{P}(\mathcal{E} > \frac{(\frac{1}{4}-\epsilon)L\alpha^{i}}{D})$
and $\Delta p = \mathbb{P}(\frac{-(\frac{1}{4}-\epsilon)L\alpha^{i}}{D} \leq \mathcal{E} \leq \frac{(\frac{1}{4}-\epsilon)L\alpha^{i}}{D})$
that $N^{+}_{i} > N_i p_1+ \frac{N_i \Delta p}{2}$ is satisfied if $w^{*}_{j} \in [x_{j},x_{j}+(1-\alpha)(y_{j}-x_{j})]$ with probability
at most $e^{-\frac{n_{i}(\Delta p)^{2}}{10}}$. Similarly, $N^{+}_{i} \leq N_i p_1+ \frac{N_i \Delta p}{2}$ is satisfied if if $w^{*}_{j} \in [y_{j},y_{j}-(1-\alpha)(y_{j}-x_{j})]$ 
with probability at most $e^{-\frac{n_{i}(\Delta p)^{2}}{10}}$.
We can use Lemma \ref{segment_lemma} since (as it is easy to notice) in the $i^{th}$ phase $\Delta p$ is exactly 
$\Delta p_{i} = \mathbb{P}(-\frac{(\frac{1}{4}-\epsilon)L\alpha^{i}}{D} \leq \mathcal{E} \leq \frac{(\frac{1}{4}-\epsilon)L\alpha^{i}}{D})$
and $p_{1}$ is exactly $\mathbb{P}(\mathcal{E} > \frac{(\frac{1}{4}-\epsilon)L\alpha^{i}}{D})$.
We obtain the following: the probability that
there exists $i$ such that $w^{*} \notin \mathcal{H}\mathcal{C}_{0} \cap \cdots \cap \mathcal{H}\mathcal{C}_{i}$ is at most:
$O(\sum_{i=0}^{s} e^{-\frac{n_{i}(\Delta p_{i})^{2}}{10}})$. Substituting in that expression the formula on $n_{i}$, 
and noticing that the number of all the phases of the algorithm is logarithmic in $T$, $D$ and $d$, we get
the bound $O(\frac{\log(dDT)}{T^{3}})$.
Thus, according to our previous remarks, we conclude that with probability at least $1 - O(\frac{\log(dDT)}{T^{3}} + \frac{d\log(dT)}{T^{30}})$
\textit{OnlineBisection} algorithm makes an average error at most: $\epsilon_{av} = O(\frac{1}{T}(D^{3}d^{\frac{3}{2}}r\log(T)h_{T}+\frac{\sqrt{d}T}{h_{T}}))$.
As mentioned before, we complete the proof by using the formula: $h_{T} = \frac{\sqrt{T}}{\log(T)}$.
\bbox

\section{Conclusions}

We presented in this paper the first $\tilde{O}(\frac{1}{\sqrt{T}})$-error algorithm for database reconstuction in the online setting, using logarithmic memory and
$O(dD)$-time per query. It is designed for the highly
challenging, yet very realistic setting, where the answers given by the database are heavily perturbated by a random noise and there
exists a strong privacy mechanism (binary oracle $\mathcal{O}$) that aims to protect the database
against an adversary attempting to compromise it. We show that even if the learning algorithm receives only binary answers on the database side and
needs to learn database-vector $w^{*}$ with high precision at the same time it is being evaluated, it can still achieve very small average error. 
We assume that the query-space 
is low-dimensional but this fact is needed only to guarantee that the term $r=\frac{2}{p_{\mathcal{D},\phi}^{2}}$ from the bound on the error is not
exponential in $D$. The low-dimensionality assumption is indispensable here if one wants to achieve average error of the order $o(1)$ in a 
nontrivial setting with random noise. 
\textit{OnlineBisection} algorithm adapts next threshold values sent to the binary oracle $\mathcal{O}$ to its previous answers in order to 
obtain good approximation of the projection of a database-vector $w^{*}$ onto a low-dimensional query-space $\mathcal{U}$.


\bibliographystyle{unsrt}
\bibliography{database-icalp}

\newpage

\appendix

\section{Analysis of the running time of the algorithm and memory usage}
\label{sec:running}

We start with the analysis of the running time of \textit{OnlineBisection}. First we will show that the linear program used by the algorithm to determine the threshold in each round has a closed form solution.

\begin{lemma} \label{lem:closedform} For any query-vector $q$, $\mathcal{I}_{1}=[x_{1},y_{1}],\ldots,\mathcal{I}_{d}=[x_{d},y_{d}]$ and orthonormal basis $\mathcal{C}=\{e^{1},\ldots,e^{d}\}$ the value
$$
\max_{f_{1} \in \mathcal{I}_{1},\ldots,f_{d} \in \mathcal{I}_{d}}\sum_{i=1}^{d} f_{i} e^{i} \cdot q
$$
if given by
$$
opt = \sum_{j \in \mathcal{J}_{+}} y_{j} e^{j} \cdot q + \sum_{j \in \mathcal{J}_{-}} x_{j} e^{j} \cdot q
$$
where $\mathcal{J}_{+} = \{i \in \{1,\ldots,d\}: e^{i} \cdot q \geq 0\}$ and $\mathcal{J}_{-} = \{i \in \{1,\ldots,d\}: e^{i} \cdot q < 0\}$.
\end{lemma}
\Proof
Take some point: $c_{1}e^{1} + \cdots + c_{d}e^{d}$, where: $x_{i} \leq c_{i} \leq y_{i}$ for $i=1,\ldots,d$. 
For $j \in \mathcal{J}_{+}$ the following is true:
$c_{j}e^{j} \cdot q \leq y_{j} e^{j} \cdot q$, since: $c_{j} \leq y_{j}$ and $e^{j} \cdot q \geq 0$.
Similarly, for $j \in \mathcal{J}_{-}$ we have: $c_{j}e^{j} \cdot q \leq x_{j} e^{j} \cdot q$, again by the definition 
of $\mathcal{J}_{-}$. Combining these inequalities we get that for every point $v$ in the hypercube 
$\mathcal{H}\mathcal{C}$ induced by $\mathcal{I}_{1},\ldots,\mathcal{I}_{d}$ and $\mathcal{C}$ the following is true:
$v \cdot q \leq opt$. Besides clearly there exists
$v^{*} \in \mathcal{H}\mathcal{C}$ such that: $v^{*} \cdot q = opt$.
\bbox

Now let us fix a query $q$. It is easy to notice that $q$ is being processed by
the algorithm in $O(dD)$ time. Indeed, a single query requires updating $O(d)$ variables of the form $N^{+}_{i}$, $N^{-}_{i}$
and computing the closed-form solution given in Lemma \ref{lem:closedform} in $O(dD)$ time. 
Computing dot product of the query with the given approximation of the database vector clearly takes $O(D)$ time.
Thus \textit{OnlineBisection} runs in the $O(dD)$-time per query.
Notice that \textit{OnlineBisection} algorithm does not store any nontrivial data structures, only segments: $\mathcal{I}_{1},\ldots,\mathcal{I}_{d}$,
counts: $N^{+}_{i}$, $N^{-}_{i}$ for $i=1,\ldots,d$ and a constant number of other variables.
The counts can be represented by $O(\log(T))$-digit numbers thus we conclude that \textit{OnlineBisection} runs in the $O(\log(T))$-memory.

\section{Proof of Lemma \ref{segment_lemma}}

\Proof
The proof follows from standard concentration inequalities. Let $\delta_{1}, \delta_{2} > 0$.
Note that $E(Z^{m}) \leq m p_{1}$ and $E(W^{m}) \geq mp_{1}+ m\Delta p$.
Denote $\mu_{2} = E(W^{m})$. Note that by Chernoff's inequality we have:
$\mathbb{P}(Z^{m} \geq (1+\delta_{1})\mu_{1}) \leq e^{-\frac{\delta_{1}^{2}}{2+\delta_{1}}\mu_{1}}$.
Similarly, $\mathbb{P}(W^{m} \leq (1-\delta_{2})\mu_{2}) \leq e^{-\frac{\delta_{2}^{2}}{2+\delta_{2}}\mu_{2}}$.
Take: $\delta_{1} = \frac{m \Delta p}{2 \mu_{1}}=\frac{\Delta p}{2 p_{1}}$, 
$\delta_{2} = \frac{m \Delta p}{2 \mu_{2}}$. 
Using these values of $\delta_{1}$ and $\delta_{2}$, we obtain:
$\mathbb{P}(Z^{m} \geq \mu_{1} + \frac{m \Delta p}{2}) \leq e^{-\frac{1}{1+\frac{2}{\delta_{1}}}\frac{m \Delta p}{2}}$. 
Similarly, $\mathbb{P}(W^{m} < \mu_{1} + \frac{m \Delta p}{2}) \leq \mathbb{P}(W^{m} \leq \mu_{2} - \frac{m \Delta p}{2}) \leq e^{-\frac{1}{1+\frac{2}{\delta_{2}}}\frac{m \Delta p}{2}}$. 
Notice that $\delta_{1},\delta_{2} \geq \frac{\Delta p}{2}$ (the latter inequality holds because obviously: $\mu_{2} \leq m$).
Thus we get: $\mathbb{P}(Z^{m} \geq \mu_{1} + \frac{m \Delta p}{2}) \leq e^{-\frac{m (\Delta p)^{2}}{2(4+\Delta p)}}$
and $\mathbb{P}(W^{m} < \mu_{1} + \frac{m \Delta p}{2}) \leq e^{-\frac{m (\Delta p)^{2}}{2(4+\Delta p)}}$.
Since $\Delta p \leq 1$, the proof is completed.
\bbox

\section{Proof of Lemma \ref{cuboid_lemma}}

\Proof
Note first that for any $d$-dimensional hypercube $\mathcal{H}\mathcal{C} \in \mathbb{R}^{D}$ of side length $l$, two vectors:
$w^{1},w^{2} \in \mathcal{H}\mathcal{C}$ and a vector $q=(q_{1},\ldots,q_{D})$ such that: $q_{i}=1$ for
$i=1,\ldots,d$ the following is true: $|w^{1} \cdot q - w^{2} \cdot q| \leq l\sqrt{dD}$. This comes from the fact that:
$\|w^{1}-w^{2}\|_{2}\leq l\sqrt{d}$, $\|q\|_{2} \leq \sqrt{D}$ and Cauchy-Schwarz inequality.
Thus we see that the cumulative error $\epsilon^{1}_{cum}$ made by the algorithm for the first $\sum_{i=0}^{s}k_{i}$ queries satisfies: 
$\epsilon^{1}_{cum} \leq \sum_{i=0}^{s} k_{i} L\alpha^{i} \sqrt{dD} \leq
L\sqrt{dD}r\sum_{i=1}^{s} m_{i} \alpha^{i}$. Therefore we have: $\epsilon^{1}_{cum} \leq CL\sqrt{dD}r\log(T) \sum_{i=0}^{s} 
\frac{\alpha^{i}}{\phi_{\epsilon}^{2}(i)}$. We can write: $\epsilon^{1}_{cum} \leq CL\sqrt{dD}r \log(T) 
\sum_{i=0}^{t} \frac{\alpha^{i}}{\phi_{\epsilon}^{2}(i)} + CL\sqrt{dD}r \log(T) \sum_{i=t+1}^{s}\frac{\alpha^{i}}{\phi_{\epsilon}^{2}(i)}$, where $t$ is the smallest index
such that $\rho(x) \geq \frac{\rho(0)}{\sqrt{2}}$ for $x \in [-\frac{\alpha^{t}}{8},\frac{\alpha^{t}}{8}]$.
Since $\rho$ is continuous at $0$, $t$ is well-defined. Notice that $t$ does not depend on $d$, $D$ and $T$, but only on the
random variable $\mathcal{E}$ and constant $\alpha$.
Observe that $CL\sqrt{dD}r \log(T) 
\sum_{i=0}^{t} \frac{\alpha^{i}}{\phi_{\epsilon}^{2}(i)} \leq CL\sqrt{dD}r \log(T) \frac{t}{\phi^{2}_{\epsilon}(t)}
\leq \frac{2CL\sqrt{d}D^{\frac{5}{2}}r \log(T)t}{\rho^{2}(0)\alpha^{2t}(\frac{1}{2}-2\epsilon)^{2}}$, where the last inequality follows immediately from the definition of $t$ (density $\rho$ on the interval considered in the definition of $\phi_{\epsilon}(t)$ is at least $\frac{\rho(0)}{\sqrt{2}}$ thus the related probability is at least: the length of that interval times $\frac{\rho(0)}{\sqrt{2}}$, i.e.: $\phi_{\epsilon}(t) \geq \frac{\rho(0)}{\sqrt{2}}\frac{L\alpha^{i}(\frac{1}{2}-2\epsilon)}{D}$). Therefore the considered expression is of the order $O(L\sqrt{d}D^{\frac{5}{2}}r \log(T))$.
Now let us focus on the expression:  $\mathcal{R} = CL\sqrt{dD}r \log(T) \sum_{i=t+1}^{s}\frac{\alpha^{i}}{\phi_{\epsilon}^{2}(i)}$. From the definition of $t$ we get:
$\mathcal{R} \leq CL\sqrt{d}D^{\frac{5}{2}}r\log(T) \Pi$, where $\Pi = \sum_{i=0}^{s} \frac{2\alpha^{i}}{\alpha^{2i}(\frac{1}{2}-2\epsilon)^{2}\rho^{2}(0)}$. Therefore $\mathcal{R} \leq \frac{32CL\sqrt{d}D^{\frac{5}{2}}r\log(T)}{\rho^{2}(0)} \sum_{i=1}^{s}\alpha^{-i}$.
Thus we have: $\mathcal{R} \leq \frac{32CL\sqrt{d}D^{\frac{5}{2}}r\log(T)}{\rho^{2}(0)} \frac{\alpha}{1-\alpha} 
((\frac{1}{\alpha})^{s+1}-1) \leq \frac{32CL\sqrt{d}D^{\frac{5}{2}}r\log(T)}{\rho^{2}(0)(1-\alpha)\alpha^{s}} $. Using the formula on $s$, we get:
$\mathcal{R} \leq \frac{32CL^{2}D^{\frac{5}{2}}dr\log(T)h(T)}{\rho^{2}(0)(1-\alpha)}$.
Combining this upper bound on $\mathcal{R}$ with the upper bound on the previous expression, we obtain:
$\epsilon^{1}_{cum}  =  O(L^{2}D^{\frac{5}{2}}dr\log(T)h(T))$.
Next let us focus on the cumulative error $\epsilon^{2}_{cum}$ made by the algorithm for the remaining $T-\sum_{i=0}^{s}k_{i}$ queries.
By the definition of $s$ we know that $l(\mathcal{H}\mathcal{C}_{s}) \leq \frac{1}{\sqrt{d}h(T)}$.
This implies that for any $w \in \mathcal{H}\mathcal{C}_{s}$ we have: $\|w-w^{*}\|_{2} \leq \frac{1}{h(T)}$.
Thus clearly for any query coming in this phase the learning algorithm makes an error at most 
$\frac{\sqrt{D}}{h(T)}$ (again, by Cauchy-Schwarz inequality) and we have at most $T$ queries in this phase.
Therefore $\epsilon^{2}_{cum} = O(\frac{\sqrt{D}}{h(T)}T)$.
That completes the entire proof.
\bbox

\section{Proof of Lemma \ref{cut_lemma}}

\Proof
Denote: $\eta = e - z$. Note that $\| \eta \|_{2} \leq 2\sin(\frac{\theta}{2})$.
Take first $y \in \mathcal{H}{C}_{l}$.
We have: $m \leq z \cdot y \leq m + \beta(M-m)$.
Thus $m + \eta \cdot y \leq e \cdot y \leq m + \beta(M-m) + \eta \cdot y$.
Define: $\tilde{m} = \min_{y \in \mathcal{H}\mathcal{C}} y \cdot e$ and $\tilde{M} =  \max_{y \in \mathcal{H}\mathcal{C}} y \cdot e$. Notice that: $|\tilde{m} - m| \leq 2\sin(\frac{\theta}{2}) L \sqrt{d}$
and $|\tilde{M} - M| \leq 2\sin(\frac{\theta}{2}) L \sqrt{d}$. This follows directly from the fact that: $\|y\|_{2} \leq L\sqrt{d}$, $\| \eta \|_{2} \leq 2\sin(\frac{\theta}{2})$
and Cauchy-Schwarz inequality. Thus we obtain: $\tilde{m}-2\sin(\frac{\theta}{2}) L \sqrt{d} + \eta \cdot y \leq
e \cdot y \leq \tilde{m} + 2\sin(\frac{\theta}{2}) L \sqrt{d} + \beta(\tilde{M}-\tilde{m} + 4\sin(\frac{\theta}{2}) L \sqrt{d}) + \eta \cdot y$.
Since, from the definition of $\tilde{M}, \tilde{m}$ and $\mathcal{H}\mathcal{C}$ we have:
$\tilde{M}-\tilde{m} = L$, we obtain: $\tilde{m}-2\sin(\frac{\theta}{2}) L \sqrt{d}+\eta \cdot y \leq e \cdot y \leq \tilde{m} + 2\sin(\frac{\theta}{2}) L \sqrt{d} + \beta(L + 4\sin(\frac{\theta}{2}) L \sqrt{d}) + \eta \cdot y$. Therefore 
$\max_{y \in \mathcal{H}\mathcal{C}_{l}} e \cdot y - \min_{y \in \mathcal{H}\mathcal{C}_{l}} e \cdot y
\leq L(\beta + 8\sin(\frac{\theta}{2}) \sqrt{d})$. This completes the proof of inequality \ref{maxmin_1}. The proof of inequality
\ref{maxmin_2} is completely analogous.

\bbox

\section{Online-to-batch conversion}
Throughout the paper we have considered the challenging online
scenario, where the algorithm both learns and is evaluated on a single
set of streaming queries. However, we note that the
\textit{OnlineBisection} algorithm also works well in the batch
setting, i.e. when there is a separate train and test phase.
We prove here Corollary \ref{batch_corollary}, that for clarity we state once more:
\begin{main_corollaries}
  Let $w_T$ denote the final hypothesis constructed by the
  \textit{OnlineBisection} algorithm after consuming $T$ queries drawn
  from an unknown distribution $\mathcal{D}$. Then the following
  inequality holds with probability at least $1 - O \Big(
  \frac{\log(dDT)}{T^3} + \frac{d \log(dT)}{T^{30}} \Big)$ for any future
  queries $q$ drawn from $\mathcal{D}$:
  \begin{equation*}
    \mathrm{E}_{q \sim \mathcal{D}} \big[ | w_T \cdot q - w^* \cdot q | \big]
    \leq \frac{\sqrt{D} \log(T)}{\sqrt{T}} \,.
  \end{equation*}
\end{main_corollaries}
\begin{proof}
  This simply follows from the fact that, as argued in the proof of
  Theorem~\ref{core_thm}, $w^* \in \mathcal{HC}_s$ with at least the
  probability indicated in the statement of this corollary. Furthermore, by
  definition of the algorithm, we have $w_T \in \mathcal{HC}_s$ and
  the length of the side of the hypercube $\mathcal{HC}_s \leq \log(T)
  / \sqrt{T}$. Thus, with at least the probability indicated, $|
  w_T \cdot q - w^* \cdot q | \leq \| w_t - w^* \|_2 \|q\|_2 \leq
  \frac{\log(T)}{T} \sqrt{D}$.
\end{proof}

\end{document}